\documentclass[english,aps,prper,reprint,showpacs,titlepage,longbibliography,floatfix]{revtex4-2}   

\usepackage{soul}
\usepackage{xcolor}
\soulregister{\textit}{1}

\definecolor{lightgray}{gray}{0.9}

\sethlcolor{lightgray}
\usepackage[utf8]{inputenc}
\usepackage[compatibility=false]{caption} 
\usepackage{booktabs} 
\usepackage{tabularx} 
\usepackage{ragged2e} 
\usepackage{caption} 
\usepackage[T1]{fontenc}	
\usepackage{geometry}
\usepackage{times}
\usepackage{hyperref}  
\hypersetup{colorlinks=true,urlcolor=blue,citecolor=blue,linkcolor=blue} 
\usepackage{array}
\usepackage{enumerate}
\usepackage{enumitem}
\usepackage{amsmath}
\usepackage{caption}
\usepackage{amssymb}
\usepackage{tikz}
\usepackage{graphicx}
\usepackage{multirow}
\usepackage{tcolorbox}
\usepackage{ragged2e}
\usepackage[utf8]{inputenc}

\usepackage[normalem]{ulem}
\usepackage{setspace} 
\usepackage{placeins} 
\usepackage{adjustbox} 
\begin{document}
\begin{titlepage}

\title{Exploring the Feasibility of Employing a Hybrid Machine Learning Method to Unpack Student Reasoning Patterns in Physics Essays}

 \author{Winter Allen}
 \affiliation{Department of Physics and Astronomy, Purdue University, 525 Northwestern Ave, West Lafayette, IN 47907, U.S.A.}

 \author{N. Sanjay Rebello}
 \affiliation{Dept. of Physics and Astronomy / Dept. of Curriculum \& Instruction, Purdue University, West Lafayette, IN 47907, U.S.A.} 


\begin{abstract}

We propose a novel clustering pipeline that combines two classic clustering algorithms to better understand student problem-solving strategies. This unsupervised machine learning method helps uncover patterns in reasoning without pre-defined labels. We applied it to essays written for an online multiple-choice quiz, the resulting clusters showed strong statistical alignment with students’ selected answers. We also report on the resulting clusters of the hybrid pipeline compared to that of K-Means \cite{macqueen1967some} and Hierarchal Density-Based Spatial Clustering of Application with Noise (HDBSCAN) \cite{mcinnes2017hdbscan} by analyzing the Scatter Plots, Silhouette Scores\cite{rousseeuw1987silhouettes}, and Davies Bouldin Index \cite{davies1979cluster}.

    \clearpage
  \end{abstract}

\maketitle
\end{titlepage}

\section{Introduction}
Problem solving is a highly valued skill that is essential for STEM students entering the workforce. According to the Next Generation Science Standards (NGSS), learning to define problems and design solutions is one of the key science and engineering practices \cite{ngss2013next}. For years, studies have shown that students in introductory STEM courses tend to use ineffective problem solving strategies. \cite{leonard1996using,mestre1997promoting, maloney2011rapid}. When asked to describe in writing their strategy to solve a problem, students' problem solving skills have been shown to improve \cite{leonard1996using}. Strategy writing, with appropriate scaffolds, helps students focus on the deeper structure of the problem \cite{dufresne1992constraining, gerace2001problem,mestre1993promoting}, starting with conceptual analysis \cite{docktor2010conceptual} of the problem, and avoiding novice unproductive strategies.

In Physics Education Research (PER), there are recent efforts to utilize machine learning tools in classrooms \cite{martin2023exploring, tschisgale2023}. Some are focused on assessment, while others seek to extract information from student data. Martin et al. \cite{martin2023exploring} use unsupervised machine learning methods to analyze emerging patterns in students' argumentation essays. Tschisgale et al. \cite{tschisgale2023} utilizes known unsupervised clustering techniques in conjunction with Computational Grounded Theory \cite{nelson2020}to assess student problem solving strategies. There are many methods that can be used to analyze student responses, including supervised \cite{hastie2009supervised}, semi-supervised \cite{zhu2005semi}, and unsupervised \cite{hastie2009unsupervised} machine learning. Supervised learning works by utilizing inputs to predict the outputs \cite{hastie2009supervised}. For this method, known, labeled data is required to train an algorithm to better predict patterns for unlabeled datasets. Semi-supervised learning works by using large amounts of unlabeled data, together with the labeled data, to build better classifiers \cite{zhu2005semi}. Unsupervised learning algorithms work to infer properties or patterns of a dataset without any labeled data. \cite{hastie2009unsupervised} 

Unsupervised learning is a viable technique for many researchers, including ourselves, to explore large textual datasets that are not measurable using qualitative methods. The ideal way to extract student processes is to read each essay from a student and qualitatively label their response. However, time needed to read and reliably extract information form large datasets of thousands of student responses is prohibitive. Therefore, we explore other methods to analyze student essays. 

To investigate the problem-solving strategies described in students' essays, researchers have been limited to qualitative analysis for small datasets (ADD CITATIONS). New machine learning methods and Large Language Models (LLMs) offer the promise of detecting the emergent themes from students' written strategy essays. For the last couple of years, we have been exploring popular machine learning clustering methods to detect patterns in students' essays. 

To investigate the emergent themes in student problem-solving strategy essays, we employ clustering -- an unsupervised learning technique. There are a multitude of clustering algorithms such as: K-Means \cite{macqueen1967some}, Density-based Spatial Clustering Applications with Noise (DBSCAN) \cite{ester1996density}, Hierarchical Density-Based Spatial Clustering of Applications with Noise (HDBSCAN) \cite{mcinnes2017hdbscan}, and many more. We have explored HDBSCAN and K-Means individually on a variety of data with much difficulty. Though they each have their own strengths and offers different benefits, we often found they were limited in the insights they provided us. Therefore, we began exploring other methods that capture the benefits of both algorithms. 

In this paper, we present a novel hybrid technique from well-established clustering algorithms to assess the problem-solving performance of students from their essay responses. This method utilizes the strengths of K-Means and HDBSCAN to construct well-defined, unique clusters on physics student problem-solving strategy essays. We then compare our novel hybrid approach with two traditional clustering methods: K-Means, HDBSCAN. 

K-Means \cite{macqueen1967some} is a tried-and-true method which most researchers are familiar with. HDBSCAN \cite{campello2015} is a relatively new clustering technique built to deal with noisy data. The hybrid approach we propose utilizes both strengths. Throughout our research, we have utilized both K-Means and HDBSCAN on various datasets and found that, for our work, they seemed to miss many of the patterns we expected. In an attempt to improve their performance, we have decided to combine the two and report on the resulting clusters. We will compare three methods: applying only K-Means, applying only HDBSCAN, and applying the hybrid K-Means/HDBSCAN method to the same set of data. We will report on the scatter plots, the silhouette scores \cite{rousseeuw1987silhouettes}, the Davies-Bouldin index \cite{davies1979cluster}. We will also explore how well the hybrid clusters align with student responses to a multiple choice quiz question. Finally, we will discuss the implications of our results for physics education research. 

\subsection{Research Questions}
 
Based on the issues stated above, our specific research questions (RQs):

\vspace{0.3cm}

\indent
\begin{minipage}{0.90\linewidth}
    \textbf{RQ1}: To what extent does combining K-Means and HDBSCAN into a hybrid similarity-density-based clustering method reveal more well-constructed and rich clusters than either algorithm alone? 
    \label{RQ1}
    
\vspace{0.3cm}
    \textbf{RQ2}: To what extent can we extract expected incorrect or correct strategies based upon students' answers to a multiple choice question? 
    \label{RQ2}
\end{minipage}

\section{Background} 

\subsection{Problem Solving}

Student problem solving has been a relevant and rich topic in PER for many years\cite{doktor2014, Tulminaro2007, van1991learning}. As researchers and educators, it is imperative to aid student understanding in solving physics problems. Students often prioritize memorizing final answers
over developing a deeper understanding of the problem solving process \cite{Tulminaro2007,dufresne1997solving}. To foster this growth, it is crucial not only to understand why students solve problems the way they do but also to help them reflect on their own problem-solving strategies, allowing them to develop as both learners and future scientists. 

Research has demonstrated expert-novice differences in problem solving due to differences in conceptual understanding \cite{caramazza1981naive} and knowledge organization \cite{eylon1984effects, larkin1980expert}. Novices often lack an understanding of the underlying principles \cite{leonard1996using}, and adopt different problem solving strategies than experts \cite{doktor2014}. In addition, unlike experts, novices rarely explain their solutions, reflect on their process, or consider alternative strategies \cite{dufresne1997solving}.

Several research-based strategies have addressed the issues above \cite{doktor2014},
however, systemic challenges persist.  Introductory STEM courses often use end-of-chapter, well-structured problems, give all needed information, and have one solution path. It is unlikely that these kinds of problems prepare students with 21st Century problem solving skills. Further, these problems are often delivered via commercial online homework systems that only
grade final answers, provide limited, if any feedback, reinforcing novice problem solving strategies
such as equation hunting \cite{doktor2014, hsu2004resource}. These systems also do not provide opportunities for students to provide arguments to justify their solution strategy, collaborate with others, evaluate other possible solutions, receive elaborated feedback, or revisit similar
problems and retrieve and reapply their skills. Unfortunately, in large enrollment courses it is extremely difficult to approach student instruction in other ways. Thus, developing problem-solving skills for the 21st Century workforce is a persistent issue in STEM education.

\subsection{K-Means}

K-Means sorts unlabeled data into different clusters by their similarities. The first step is to choose the number of clusters that you would like to have and initialize them at random coordinates. Then, you begin categorizing your data points to their closest cluster by calculating the Euclidean distance of the item with each of the clusters. You assign the point to the cluster and then update the cluster by shifting it to the average of data points it includes. You will iterate until an equilibrium is reached.

From MacQueen \cite{macqueen1967some}, the mathematical basis of the K-Means algorithm revolves around the concept of partitioning an N-dimensional population into $k$ sets based on a sample, aiming to minimize the within-class variance. The primary goal of the K-Means process is to find a partition $S = {S_1, S_2, ..., S_k}$ of $E^N$ (N-dimensional Euclidean space) such that the within-class variance, denoted as $W^2(S)$, is low. This within-class variance is defined as: 

$$W^2(S) = \sum_{i=1}^{k} \int_{S_i} |z - u_i|^2 dp(z) $$

where: $p$ is the probability mass function for the population. $u_i$ is the conditional mean of $p$ over the set $S_i$.

Given a $k$-tuple of points (vectors) $X = (x_1, x_2, ..., x_k)$ in $E^N$, a minimum distance partition $S(x) = {S_1(x), S_2(x), ..., S_k(x)}$ is defined as $T_i(x) = {z \in E^N: |z - x_i| < |z - x_j|, j = 1, 2, ..., k}$. This set $T_i(x)$ contains all points in $E^N$ that are strictly closer to $x_i$ than to any other $x_j$.

The partition sets are then defined as: $S_1(x) = T_1(x)$ , $S_2(x) = T_2(x) \setminus S_1(x)$, $S_k(x) = T_k(x) \setminus (S_1(x) \cup S_2(x) \cup ... \cup S_{k-1}(x))$. This convention ensures that each point in $E^N$ is assigned to the set of the nearest mean, with ties being broken by assigning to the set with the lower index.

The algorithm starts with $k$ initial means. Then, each data point is assigned to the cluster whose mean is nearest to it, according to the minimum distance partition. Next, the mean of each cluster is recalculated as the average of all data points assigned to that cluster. More formally, if we have a sequence of random points $z_1, z_2, ...$ in $E^N$ drawn independently from a distribution $p$, and a $k$-tuple of means $x^n = (x_1^n, x_2^n, ..., x_k^n)$ at iteration $n$ with associated weights $w^n = (w_1^n, w_2^n, ..., w_k^n)$, the update rule when a new point $z_{k+n}$ falls into the $i$-th cluster $S_i(x^n)$ is:$x_i^{n+1} = (w_i^n x_i^n + z_{k+n}) / (w_i^n + 1)$, $w_i^{n+1} = w_i^n + 1$, $x_j^{n+1} = x_j^n$ and $w_j^{n+1} = w_j^n$ for $j \neq i$. Initially, $x_i^1 = z_i$ and $w_i^1 = 1$ for $i = 1, 2, ..., k$.

Overall, the mathematics behind K-Means involves defining an objective function based on within-class variance, using minimum distance partitions to assign data points to clusters, and iteratively updating the cluster centers to minimize this variance. 

\subsection{HDBSCAN}

HDBSCAN is a density-based clustering algorithm based on the noise-based spatial clustering application (DBSCAN) \cite{ester1996density}. It can identify clusters of varying shapes and densities within a dataset and does not require the number of clusters to be specified beforehand. The key difference between the two is that DBSCAN requires a fixed density threshold, while HDBSCAN works well with data points of varying densities. HDBSCAN begins by building a minimum spanning tree (MST) from the distance graph. Each point in the data set is set to be a node in the graph, where the edges represent the distance. The graph is used to construct the MST. Then they extend the MST by adding an edge for each vertex where the core distance of the corresponding object is used as a weight. The extended MST is used to construct a hierarchy as a dendrogram to determine the varying densities. It starts farthest out and progressively removes the edges in decreasing order of weights. After each removal, HDBSCAN assigns labels to the connected components that contain the end vertices of the removed edges. This is to obtain the next hierarchical level by assigning a new cluster label to a component if it still has at least one edge or, if not, assigning it to be noise. The resulting tree structure represents the various clusters at differing density levels. The higher the cluster, the more dense and robust it is. 

The mathematical basis of HDBSCAN \cite{campello2015} involves several key concepts drawn from density estimation, graph theory, and hierarchical clustering principles. At its core, HDBSCAN builds upon the idea of density-based clustering and extends it to create a hierarchical structure.
The main mathematical components start with density estimation. HDBSCAN operates under the fundamental idea that a dataset is a sample from an unknown probability density function (PDF). While it doesn't explicitly construct a full PDF estimate, it uses a nonparametric approach where "the data are allowed to speak for themselves" in determining density. The algorithm implicitly uses a form of K-Nearest Neighbors (K-NN) density estimation, where the density around a point is related to the distance to its $mpts$-nearest neighbor. 

HDBSCAN follows Hartigan's \cite{Hartigan01031987} model of density-contour clusters and trees. According to this model, a density-contour cluster at a given density level $\lambda$ is a maximal connected subset $C$ of the data space where every point $x \in C$ satisfies $f(x) \geq \lambda$, and $f(x)$ is the density function. The density-contour tree is the hierarchy of nested clusters formed by varying the density threshold $\lambda$.

For a given number of minimum points $mpts$, the core distance ($d_{core}(x_p)$) of an object $x_p$ is defined as the distance from $x_p$ to its $mpts$-nearest neighbor (including $x_p$). Mathematically, if $N_\epsilon(x_p)$ is the $\epsilon$-neighborhood of $x_p$ (all points within distance $\epsilon$ of $x_p$), then $d_{core}(x_p)$ is the minimum $\epsilon$ such that $|N_\epsilon(x_p)| \geq mpts$.

The mutual reachability distance ($d_{mreach}(x_p, x_q)$) between two objects $x_p$ and $x_q$ with respect to $mpts$ is defined as the maximum of their core distances and the actual distance between them: $d_{mreach}(x_p, x_q) = \max{d_{core}(x_p), d_{core}(x_q), d(x_p, x_q)}$. This definition ensures a symmetric measure of proximity. Notably, $d_{mreach}(x_p, x_q)$ is the minimum radius $\epsilon$ such that $x_p$ and $x_q$ are $\epsilon$-reachable based on core objects (as defined in DBSCAN).

Conceptually, a complete mutual reachability graph ($G_{mpts}$) is formed where each data object is a vertex, and the weight of an edge between two objects is their mutual reachability distance.

A crucial mathematical insight is the relationship between a variant of DBSCAN focusing on core objects and Single-Linkage hierarchical clustering. Clustering at a given $\epsilon$ is equivalent to taking the Single-Linkage dendrogram built on the mutual reachability distances, cutting it at level $\epsilon$, and treating singletons with a core distance greater than $\epsilon$ as noise. This connection allows HDBSCAN to leverage the hierarchical nature of Single-Linkage in the transformed space of mutual reachability distances.

To efficiently compute the hierarchical clustering, HDBSCAN constructs a Minimum Spanning Tree (MST) of the Mutual Reachability Graph. To account for objects becoming noise at different density levels (related to their core distances), the MST is extended to MSText by adding a "self-loop" to each vertex with a weight equal to its core distance.

The HDBSCAN hierarchy is extracted from the MSText by removing edges in decreasing order of weight (mutual reachability distances and core distances). The sequence of connected components formed during this process represents the nested clusters at different density thresholds.

To obtain a more interpretable cluster tree, the HDBSCAN hierarchy is simplified by focusing on levels where "true" cluster splits occur or where clusters disappear entirely. Spurious subcomponents (smaller than a minimum cluster size, $mclSize$, which can be set to $mpts$) resulting from noise separation are not considered true splits.

To extract a flat clustering from the hierarchy, a measure of cluster stability $S(C_i)$ is defined for each cluster $C_i$ in the simplified tree. This stability measure is related to the "excess of mass" of the cluster, conceptually derived from the integral of the difference between the maximum density at which a point $x$ belongs to $C_i$ and the minimum density of $C_i$. For a finite dataset, this is adapted by considering the range of density thresholds (represented by $\epsilon$ values) over which the cluster persists and its members remain within it.

The problem of extracting the most "prominent" non-overlapping clusters is formulated as an optimization problem: $$ \max_{\delta_2, \ldots, \delta_\kappa} J = \sum_{i=2}^\kappa \delta_i S(C_i) $$ subject to $\delta_i \in {0, 1}$ and exactly one $\delta(\cdot) = 1$ in each path from a leaf cluster to the root. Here, $\delta_i = 1$ if cluster $C_i$ is included in the flat solution, and $\delta_i = 0$ otherwise. A dynamic programming-like approach is used to efficiently find the optimal set of clusters by calculating a value $\hat{S}(C_i)$ for each node in the simplified cluster tree, representing the maximum achievable stability in the subtree rooted at $C_i$.

A novel outlier detection measure is derived from the HDBSCAN hierarchy. The GLOSH (Global-Local Outlier Scores from Hierarchies) score for an object $x_i$ is calculated as: $$ GLOSH(x_i) = 1 - \frac{\epsilon_{max}(x_i)}{\epsilon(x_i)} $$ where $\epsilon(x_i)$ is the lowest radius at which $x_i$ is no longer considered noise (i.e., it belongs to a cluster), and $\epsilon_{max}(x_i)$ is the lowest radius at which the cluster $x_i$ belonged to (or any of its superclusters in the hierarchy) disappears entirely (all its objects become noise). This score aims to unify global and local perspectives of outlierness.
By combining these mathematical concepts, HDBSCAN provides a comprehensive framework for density-based cluster analysis, visualization, and outlier detection.

\subsection{Novel Hybrid Method}

The hybrid method combines the strengths of K-Means and HDBSCAN. We start by applying K-Means clustering to the vectorized data. Recall, the basic mathematical formulation of K-Means involves cluster assignment and centroid updating. To assign each point to the nearest clusters: $$c_i=\arg \max_{k} || x_i-\mu_k||^2$$

To update each cluster center by taking the mean of all the points in the cluster: $$\mu_k=\frac{1}{N_k}\sum_{i:c_i=k}x_i$$ This is repeated until the centroids reach an equilibrium.

A similarity matrix, $S_{ij}$ is then constructed to capture points that belong to the same cluster; however this does not account for distances. To ensure points in the same cluster remain similar and points far apart have lower similarity values we employ a smoothing process computing a pairwise Euclidean matrix $D_{ij}$. $$D_{ij}=|| x_i-c_j||_2$$ 

It is then normalized by dividing by the maximum distance: $$D_{ij}^{norm}=\frac{D_{ij}}{max(D)}$$

The exponential weighting function is then applied: $$S'_{ij}=e^{-D_{ij}^{norm}}S_{ij}$$

We transform this into distance by subtracting $S'_{ij}$ from $1$: $$D'_{ij}=1-S'_{ij}$$

To apply this into HDBSCAN we begin with calculating the core distance to the $m_{pts}$-th nearest neighbor, as normal. We then use $D'_{ij}$ from K-Means to calculate the mutual reachability distance: $$d_{mreach}(x_i,x_j)=max(d_{core}(x_i),d_{core}(x_j),D'_{ij})$$

The algorithm then continues on to build an MST as previously outlined and subsequent steps.

\section{Methodology}

\subsection{Dataset}
The dataset consists of student responses to a bonus strategy question on weekly quizzes in a first-semester calculus-based physics class for future engineers, at a large U.S. Midwestern land grant University. The  enrollment of the course in the spring 2024 semester in which the data were collected is N = 1447. 

The course is built around three key principles: momentum, energy, and angular momentum, and it follows Chabay and Sherwood's \textit{Matter \& Interactions} \cite{chabay2011matter}. The responses are from Quiz 06 of the Spring 2024 semester, which falls in the energy principle portion of the course. The quiz is multiple choice with five versions randomly distributed to the students. The strategy question students were asked before they received the multiple-choice question is: 

\vspace{0.3 cm}
\textit{You throw a ball of known mass in the air vertically upward. The ball leaves your hand at a known speed and reaches a known maximum height above your hand. Assume that the work done by air drag is the same on the way down as it is on the way up. Describe in WORDS the strategy you would use to find the final speed of the ball when it falls back to your hand. Your response should be at least 50 words long.  Do NOT type any symbols, formulae, equations or numbers in your answer. Your answer must strictly use WORDS only.} 

\vspace{0.3 cm}

This strategy questioned corresponded to the following multiple-choice question:

\vspace{0.3 cm}
\textit{You throw a ball of mass \textbf{$M$} $kg$ in the air vertically upward. The ball leaves your hand at a speed of \textbf{$v_{i}$} $m/s$ and reaches a maximum height of \textbf{$H_{max}$} $m$ above your hand. What is the final speed of the ball as it falls back to your hand? Assume that the work done by air drag is the same on the way down as it is on the way up. Magnitude of acceleration due to gravity is \textbf{$10.0$} $m/s^2$. Select the closest answer.}
\vspace{0.3 cm}

The only variations between versions are the numerical values of mass of the ball, the initial speed of the ball, and the maximum height. Students are given the option of five multiple choice answers. One answer corresponds to the correct answer, which is expected to have very few correct strategies. The majority of the four other choices correspond to specific incorrect strategies that students may have taken. The choices correspond to five different formulas and potential strategies, shown in Table \ref{table:multichoice}. 

\begin{table}[htbp]
\caption{Multiple Choice Answers}
\begin{ruledtabular}
\begin{tabular}{ccp{4cm}} 
\textbf{Answer} & \textbf{Formula} & \textbf{Probable Strategy} \\ 
\hline
\textbf{Correct} & $\sqrt{4 g h - v_i^2}$ & Breaking the problem into two parts (ball going up and ball going down) and applying the energy principle. \\
\\
\textbf{Wrong 1} & $v_i$ & Assuming drag force plays no role and the final velocity will be the same as the initial. \\
\\
\textbf{Wrong 2} & $\sqrt{v_i^2-2 g h}$ & Did not break the problem into two parts, ignoring drag force, then applied the energy principle by treating final potential energy as $m g h$. \\
\\
\textbf{Wrong 3} & $\sqrt{v_i^2-g h}$ & Breaking the problem into two parts (ball going up and ball going down) and applying the energy principle. \\
\\
\textbf{Wrong 4} & $v_i D^2$ & $D$ represents an arbitrary fraction of $v_i$. Drag force has an effect but students calculate a trivial proportion. \\
\end{tabular}
\label{table:multichoice}
\end{ruledtabular}
\end{table}

The wrong answers 2 and 3 can be attributed to conceptual misunderstandings or algebraic mistakes. However, the wrong answers 1 and 4 likely lie in fundamental misunderstandings.

\subsection{Data Cleaning and Vectorization}
To prepare textual data for analysis, we applied a series of pre-processing steps using the Natural Language Toolkit (NLTK) \cite{loper2002nltk}. We used the English stopword list from NLTK to remove common words that do not contribute to meaning, such as "the", "of", "a" , etc. Lemmatization was performed using WordNetLemmatizer from NLTK, which reduces words to their base form (e.g., "launched" to "launch") while preserving their grammatical meaning. We employed the SpellChecker from the pyspellchecker \cite{barrus2019pure} library to detect and correct misspelled words. If a word was identified as misspelled, it was replaced with its most likely correct form. Then the text was converted to lowercase to maintain consistency.

To tokenize the text we used the word tokenizer feature from NLTK, splitting it into individual words. Non-alphabetic characters and punctuation were removed to retain only meaningful tokens. We then filtered responses with fewer than 10 words to ensure that only substantive textual data was retained for further analysis. Any responses smaller than that will likely be trivial or primarily empty. Before cleaning, there 1408 student responses. After cleaning, there were 1018 responses with an average length of 39.13 words. 

We used the Term Frequency-Inverse Document Frequency (TF-IDF) transformation, TfidfVectorizer from Scikit-learn \cite{pedregosa2011scikit}, to convert preprocessed text into a weighted numerical representation. To optimize feature extraction, we used the default parameters. This resulted in ignoring terms that appear in more than 95 \% of documents and ensuring hat only words appearing in at least 2 documents are considered. We included an optional step to reduce the high-dimensional TF-IDF feature space while preserving important information, we applied Singular Value Decomposition (SVD) using TruncatedSVD from sklearn.decomposition. The dimensionality was reduced to 100 components, balancing computational efficiency with information retention. A random seed was set to ensure reproducibility. This transformation aids in improving clustering performance by removing noise and reducing sparsity in the feature representation.

\subsection{K-Means Clustering}

To apply the K-Means clustering algorithm, we used the Scikit-learn \cite{pedregosa2011scikit} library. This method partitions the data into k clusters based on feature similarity. We determine the optimal number of clusters based upon the number of multiple choice responses. There are 5 possible answers for students to choose, so we set the number of clusters needed to be k=5. The clustering was performed on a TF-IDF feature matrix, which represents the text data in a high-dimensional space. A random seed  was used for reproducibility, and the number of initializations was set to ensure a robust clustering outcome. The fit predict method was used to fit the K-Means model to the TF-IDF matrix and assign cluster labels to each document

\subsection{HDBSCAN Clustering}

We employed the HDBSCAN algorithm from hdbscan library \cite{mcinnes2017hdbscan} to identify natural groupings in the data. We set the parameters to attempt to get 5 resulting clusters. or as close as possible, to match with the number of multiple-choice options. The model was trained using TF-IDF-reduced features, and each document was assigned a cluster label using the fit predict feature.

\subsection{Hybrid Clustering}

To refine clustering results, we implemented a hybrid clustering approach that first applies K-Means clustering to construct a similarity matrix and then utilizes HDBSCAN to extract more meaningful clusters from the structured distance representation.

\begin{figure}
    \centering
    \includegraphics[width=1\linewidth]{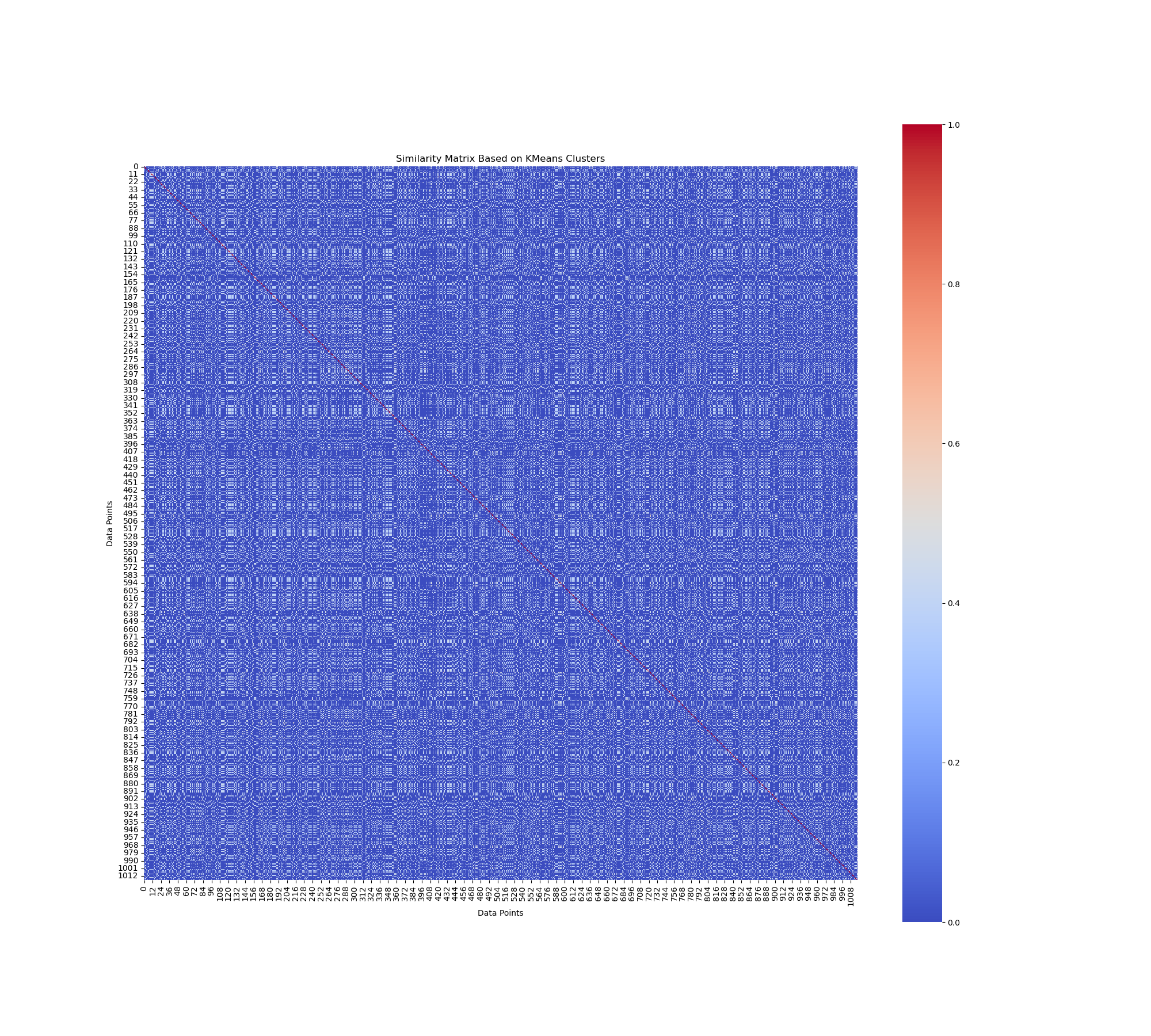}
    \caption{Similarity Matrix}
    \label{fig:matrix}
\end{figure}

We initially applied K-Means clustering using Scikit-learn \cite{pedregosa2011scikit} to segment the data into k=5 clusters. The clustering was performed on the TF-IDF feature matrix, and the assigned cluster labels were extracted for each data point. To capture document relationships beyond strict K-Means assignments, we built a similarity matrix. Since we had 1018 responses after pre-processing, we ended up with a 1018 by 1018 matrix, Figure \ref{fig:matrix}.  Where an identical response has a value of 1 (red) and responses completely dissimilar have a value of 0 (dark blue). Values between 0 and 1 emerge between responses that are in the same clusters but have different characteristics, or they are in different clusters but share similar features.

This matrix was further smoothed using pairwise Euclidean distances from Scikit-learn \cite{pedregosa2011scikit} to weight the similarities exponentially:

$$similarity(i,j)=e^{-\frac{distance(i,j)}{max distance}} \times binary similarity (i,j)$$

Finally, recall,  the distance matrix required for HDBSCAN was derived as:

$$distancematrix= 1 - similarity matrix$$

This transformation ensured that the model correctly interpreted close relationships as low distances. There is a direct correlation between similarity and distance when looking at the dimensional space of the clusters. The farther away the points are, the more dissimilar they are. Therefore, a similarity of 0.4 would be converted to 0.6. Responses compared and having a value of one would have a distance of zero as they are exactly the same and, therefore, would have no difference in distance.

Using the distance matrix constructed from the similarity matrix of the K-means clusters, we ran HDBSCAN \cite{mcinnes2017hdbscan} to extract more refined clusters. We utilized the precomputed distance matrix instead of raw feature vectors. Finally, the fit predict method assigned the final cluster labels based on the distance matrix

\subsection{Visualization and Evaluation}

To visually inspect the clustering results, we applied t-Distributed Stochastic Neighbor Embedding (t-SNE) \cite{van2008visualizing} from Scikit-learn to the data set. We then plotted using Matplotlib \cite{hunter2007matplotlib}. To understand and interpret each cluster, we identified the most important words per cluster using TF-IDF feature importance. For each cluster, the average TF-IDF value was computed across all documents in that cluster, and the top 10 words with the highest TF-IDF scores were selected as representative words for each cluster. To provide qualitative insights into each cluster, we also identified a representative response per cluster. The centroid of each cluster was computed as the mean of all document vectors in the cluster. The response closest to the centroid was selected.

To assess the quality of the clusters, we computed two unsupervised evaluation metrics: Silhouette Score \cite{rousseeuw1987silhouettes} and Davies-Bouldin Index \cite{davies1979cluster}. The silhouette score measures how well-separated the clusters are, with a score close to 1 indicating more distinct clusters. Davies-Bouldin Index evaluates cluster compactness and separation, where lower values indicate better clustering. We calculated these metrics (excluding noise when appropriate) to compare the different clustering algorithms.

\section{Results and Discussion}

To answer the first research question, we compared the results of the three clustering methods to the same set of student problem-solving strategy essays. We compare the scatter plots, silhouette scores (SSs), and Davies-Bouldin Indexes (DBIs). To answer the second research question, we use the top ten words and representative essay of each resulting cluster of the hybrid method to assess if responses of a certain cluster correspond to the multiple choice answer selected by the student. We report on the statistical significance of a response in a certain cluster compared to the selected answer.

\subsection{Cluster Comparison}
To compare the hybrid method performance to standard clustering algorithms, we evaluated the scatter plots, SSs, and DBIs. The results of the evaulation metrics are in Table \ref{table:scores}. For silhouette scores, the closer to 1 the clusters achieve the more distinct and unique they are. For DBI, the smaller the number the better the resulting clusters are. K-Means seems to perform the worst, in terms of cluster separation. The SS is the lowest, while its DBI is the highest. HDBSCAN (excluding noise) performs slightly better, but still the scores are fairly poor. Finally, the hybrid method outperforms both methods substantially, with the highest SS and lowest DBI.

Another important note is the way the clusters are distributed, shown in Table \ref{table:dist}. HDBSCAN throws a substantial amount of responses to noise, while this could be mitigated by adjusting parameters, to get as close to five clusters as possible it was necessary.

\begin{figure}
    \centering
    \includegraphics[width=1\linewidth]{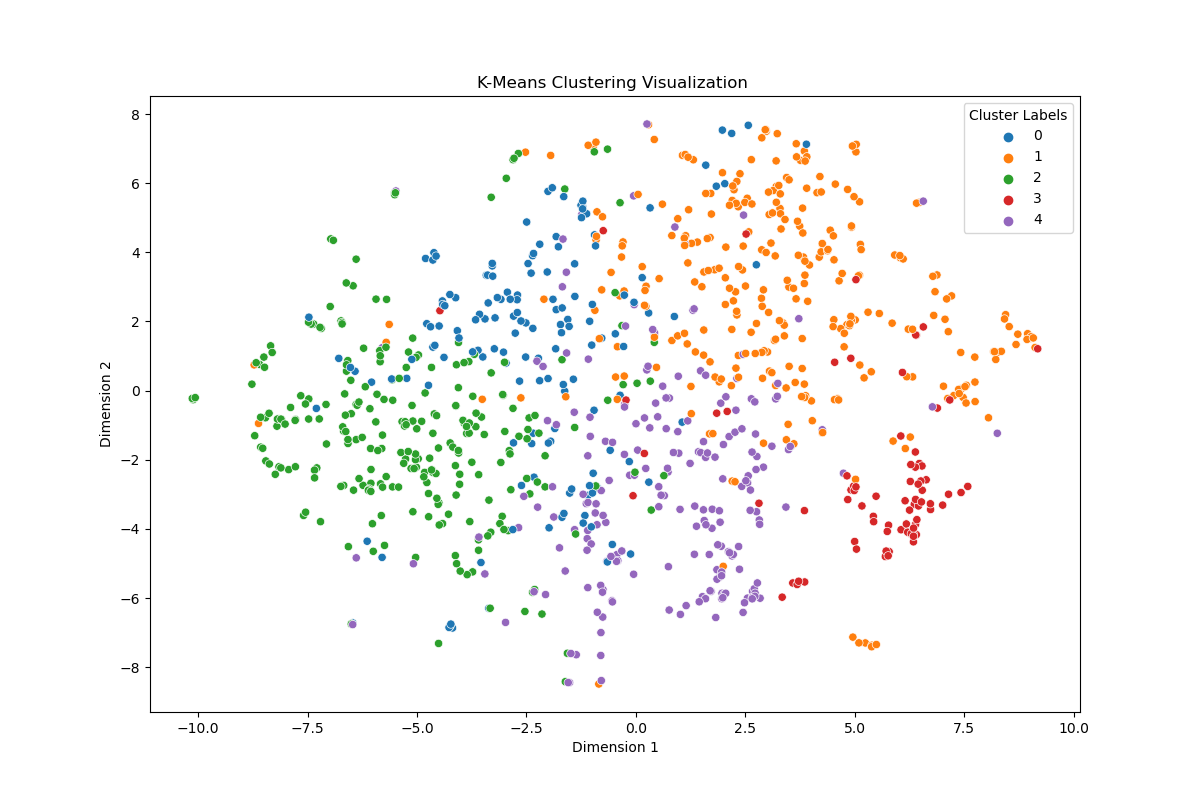}
    \caption{Scatter Plot for K-Means Clusters}
    \label{fig:kmeans_scatter}
\end{figure}

\begin{figure}
    \centering
    \includegraphics[width=1\linewidth]{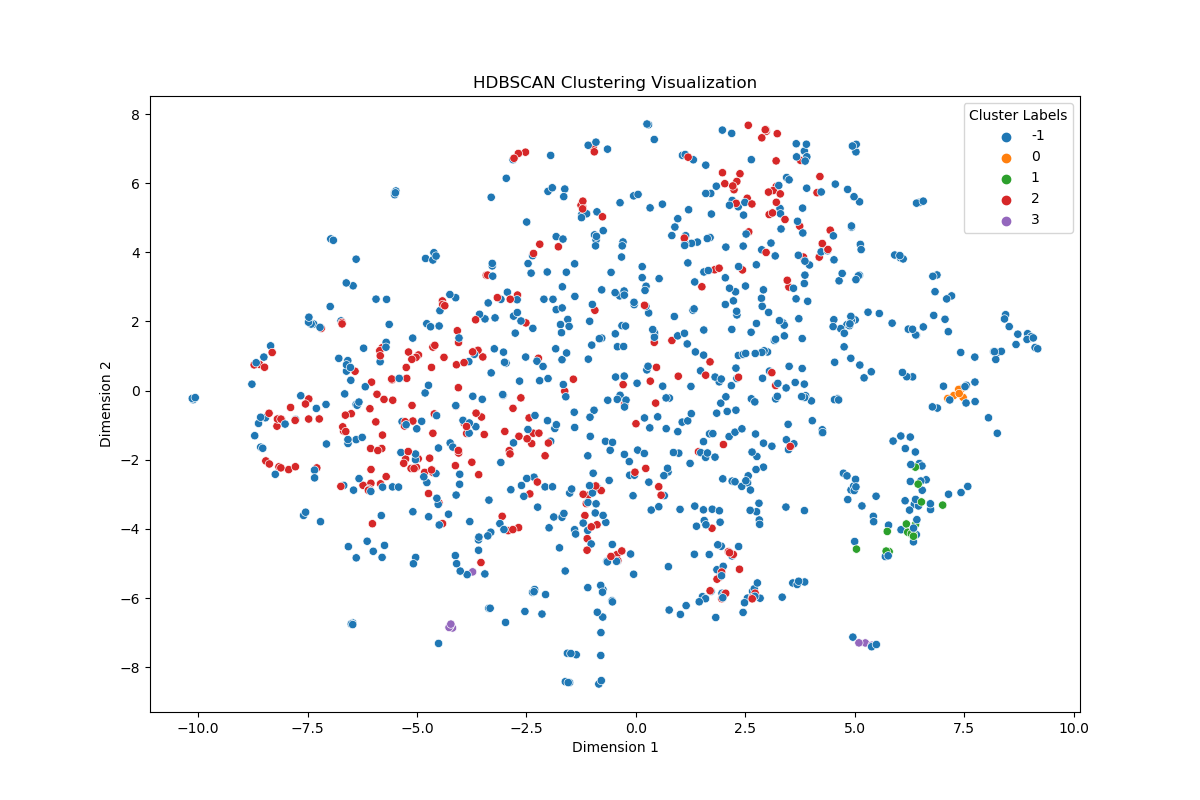}
    \caption{Scatter Plot for HDBSCAN Clusters}
    \label{fig:hdbscan_scatter}
\end{figure}

\begin{figure}
    \centering
    \includegraphics[width=1\linewidth]{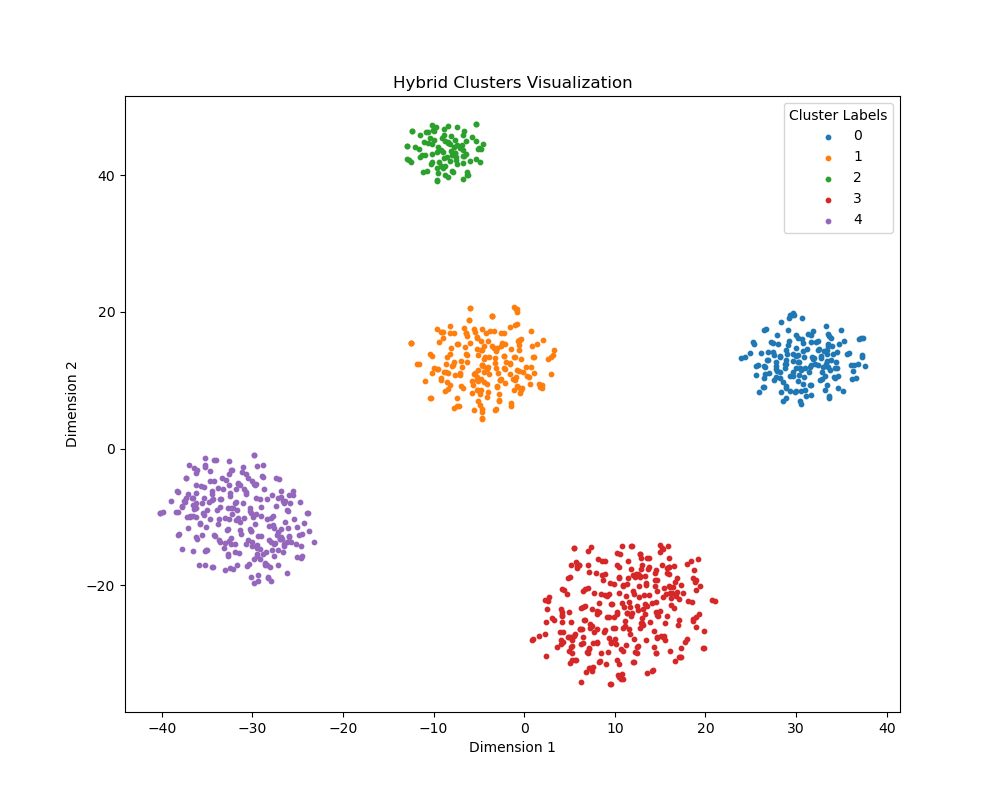}
    \caption{Scatter Plot for HYBRID Clusters}
    \label{fig:hybrid_scatter}
\end{figure}

The scores we calculated are consistent with the resulting scatter plots. For the K-Means method, in Figure \ref{fig:kmeans_scatter}, visually there is significant overlap. For the HDBSCAN method, in Figure \ref{fig:hdbscan_scatter}, excluding the noise, there is not much overlap between clusters, but the clusters are not very tight or uniform in their construction. Finally, for the hybrid method, in Figure \ref{fig:hybrid_scatter}, the clusters are distinct and uniform in their construction. One question that could be raised is why the scores are not better. This is a 2D plot of high dimensional data. When looking at a 3D plot, we see more overlap between the clusters. 

\begin{table}[htbp]
\caption{Evaluation Metrics}
\begin{ruledtabular}
\begin{tabular}{cccp{4cm}} 
\textbf{Clustering Method} & \textbf{SS} & \textbf{DBI} \\ 
\hline
\textbf{K-Means} & 0.045 & 3.844 \\
\\
\textbf{HDBSCAN} & 0.087 & 1.808 \\
\\
\textbf{Hybrid} & 0.417 & 0.172 \\
\end{tabular}
\label{table:scores}
\end{ruledtabular}
\end{table}

\begin{table}[htbp]
\caption{Cluster Distribution}
\begin{ruledtabular}
\begin{tabular}{cccc} 
\textbf{Cluster} & \textbf{K-Means} & \textbf{HDBSCAN} & \textbf{Hybrid} \\ 
\hline
\textbf{0} & 152 & 6 &165\\
\\
\textbf{1} & 309 & 17 & 202\\
\\
\textbf{2} & 266 & 248 & 96\\
\\
\textbf{3} & 80 & 8 & 304\\
\\
\textbf{4} & 211 & N/A & 251\\
\end{tabular}
\label{table:dist}
\end{ruledtabular}
\end{table}

\subsection{Hybrid Cluster compared to multiple-choice answers}
We extracted the top 10 words and the representative essay of each cluster, shown in Table \ref{table:hybrid_results}. Looking at the representative response of each cluster, we were able to draw connections between the cluster and the corresponding answer. Cluster 0 corresponds with Wrong Answer 2, as when you work on the problem as described, you end up with $\sqrt{v_i^2-2gh}$. Cluster 1 corresponds to the correct answer, as this is what we would expect from a correct strategy. Cluster 2 could correspond with a correct answer depending on interpretation. Cluster 3 corresponds to Wrong Answer 1, as it states the initial and final velocities will be equal. Finally, cluster 4 is hard to label as well. It likely corresponds most to the wrong answers of 2 or 3, since no air drag was discussed. However, it depends on the algebra and assumptions the student made.

\begin{table*}[htbp]
\caption{Hybrid Cluster}
\begin{ruledtabular}
\begin{tabular}{>{\centering\arraybackslash}c p{4cm} p{8cm}} 
\textbf{Cluster}  & \textbf{Top Words}& \textbf{Representative Essay} \\ 
\hline
\textbf{0}   & energy, potential, kinetic, ball, final, height, speed, velocity, hand, equal & I would use the conservation of energy principle to determine the final speed of the ball when it reaches my hand after reaching a known maximum height. I would take all of the energy components of the inital position and set it equal to the energy components of the final position, then use that equation to solve for the final velocity of the ball when it falls back into my hand.\\
\\
\textbf{1}  & energy, work, kinetic, drag, air, potential, ball, final, way, initial & To find the final speed of the ball when it falls back down into my hand I would take the mass of the ball and multiply it with gravity and height because when the ball reaches its maximum the inital velocity is zero and then I would find the air resentence that acts on it and subtract it from the previous calculation. This should get me the final speed of the ball. \\
\\
\textbf{2}  & time, velocity, mass, height, gravity, squared, final, ball, initial, half & I would use the energy principle to find the final speed of the ball. We can find the max height of the ball and use the acceleration due to gravity to find the final speed by subtracting drag. \\
\\
\textbf{3}  & speed, ball, force, air, final, way, velocity, hand, drag, gravity & The strategy that I would use to find the final speed of the ball when it falls back to my hand is simple. I would assume that the force of drag cancels out the initial velocity meaning that the final would be the same. THe initial and the kinetic energies in this situation would be equal for that cancellation to occur, so it would mean that the initial and final velocities is the same.\\
\\
\textbf{4}  & energy, work, change, kinetic, final, equal, velocity, ball, solve, set & By using the kinetic energy formula, we can find the final velocity by setting it equal to the work done. The work can be found by multiplying the mass, gravity, and the height. By setting the initial speeds and the initial components equal to the final components we can find the final speed.\\
\end{tabular}
\label{table:hybrid_results}
\end{ruledtabular}
\end{table*}

\begin{figure}
    \centering
    \includegraphics[width=1\linewidth]{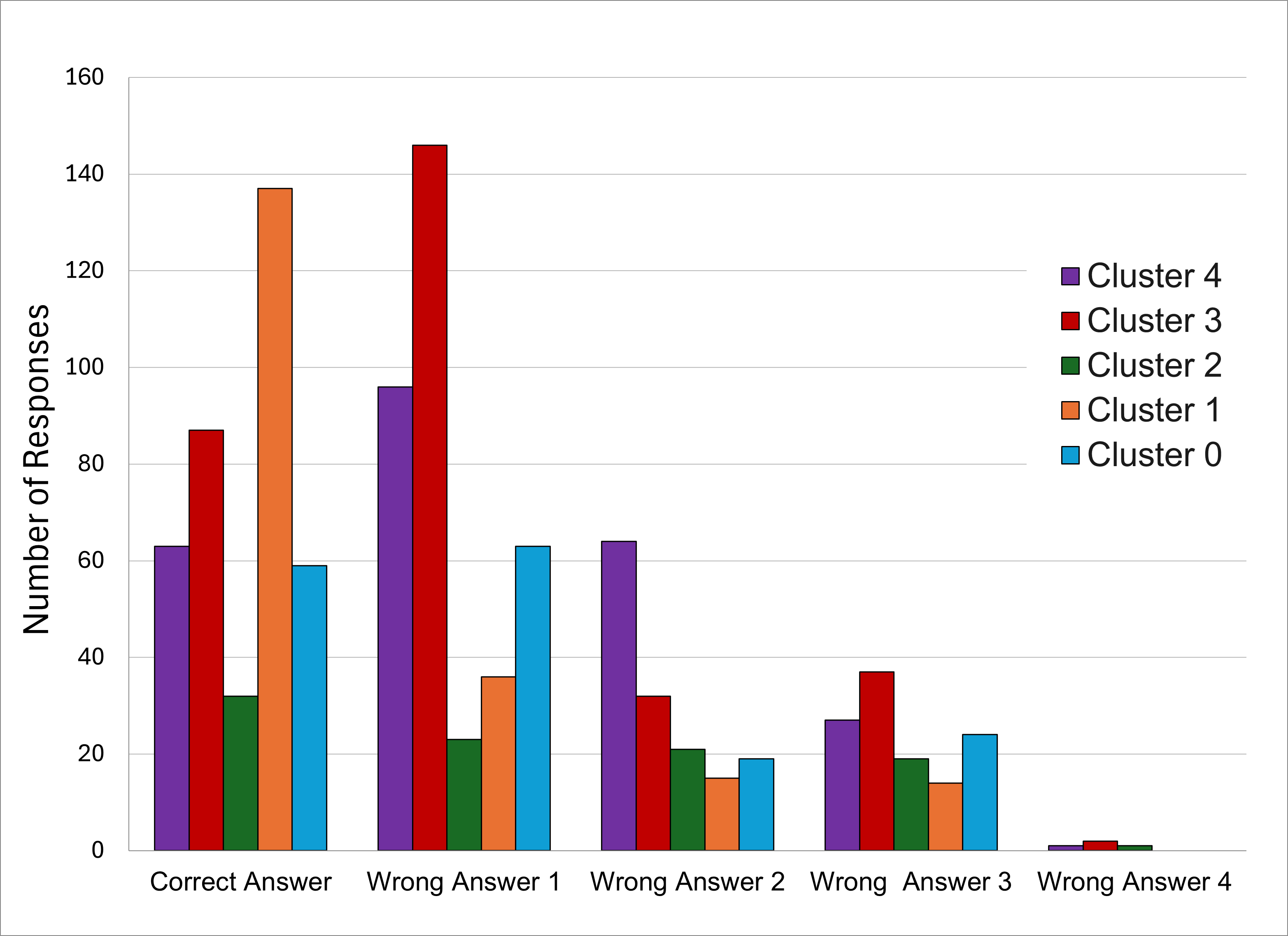}
    \caption{Distribution of HYBRID Clusters in relation to answers chosen}
    \label{fig:distribution}
\end{figure}

The distribution of students in each cluster and the choice they chose are shown in Figure \ref{fig:distribution}. Of the students who chose the correct answer, the most were clustered into Cluster 1, which we predicted would correspond to a correct cluster. Of the students who chose Wrong Answer 1, the most were clustered into Cluster 3, which we predicted would correspond to students who simply chose final velocity to equal initial. Of the students who chose Wrong Answer 2, the most were clustered into Cluster 4.  Wrong answers 3 and 4 were chosen by fewer people than the other three choices. This leads to it being an unreliable test of comparison between clusters. 

\begin{table*}[htbp]
\caption{Resulting Statistics}
\begin{ruledtabular}
\begin{tabular}{ccc} 
\textbf{Metric} & \textbf{Value} & \textbf{Meaning}\\ 
\hline
\textbf{$\chi^2$} & 188.825 & Large difference between observed and expected\\
\\
\textbf{p-value} & $1.422 \times 10^{-31}$ & Highly Significant \\
\\
\textbf{Cramer's V} & 0.215 & Medium Effect \\
\end{tabular}
\label{table:stats}
\end{ruledtabular}
\end{table*}

\begin{table*}[htbp]
\caption{Standardized Residuals}
\begin{ruledtabular}
\begin{tabular}{cccccc} 
\ & \textbf{Correct} & \textbf{Wrong 1} & \textbf{Wrong 2} & \textbf{Wrong 3}& \textbf{Wrong 4}\\ 
\hline
\textbf{Cluster 4} & \textbf{-3.13} & 0.66 & \textbf{4.39} & -0.52 & 0.01\\
\\
\textbf{Cluster 3} & \textbf{-2.44} & \textbf{7.37} & 1.18 & 0.14 & 0.74\\
\\
\textbf{Cluster 2} & -0.61 & -1.93 & 1.25 & \textbf{2.25} & 1.01\\
\\
\textbf{Cluster 1} & \textbf{7.16} & \textbf{-4.26} & \textbf{-2.73} & \textbf{-2.04} & -0.89\\
\\
\textbf{Cluster 0} & -0.29 & 0.52 & -1.11 & 0.99 & -0.81\\
\end{tabular}
\label{table:stand_resid}
\end{ruledtabular}
\end{table*}

To test for statistical significance we analyzed different metrics, shown in Table \ref{table:stats}. We found a $\chi^2$ of 188.825 showing a large difference between the expected and observed values, a p-value of $1.42 \times 10^{-31}$ indicating a strong statistical significance, and a Cramer's V of 0.215 indicating a medium effect. 

We also examined the standardized residuals, shown in Table \ref{table:stand_resid}. For standardized residuals, we  compare our observed values with what we would expect of clustering was completely independent of the selected answer. A value above 2 or less than -2 indicates an emerging pattern. The significant deviation between observed and expected answer distributions ($\chi^2$ = 188.82, p < .001) indicates that student answer choice was not independent of cluster membership. Standardized residuals revealed that Cluster 1 selected the correct answer (residual = +7.16) far more often than expected, while selecting all distractors less than expected (e.g., Wrong Answer 1 residual = -4.26). Cluster 3 heavily favored Wrong Answer 1 (residual = +7.37) and under-selected the correct answer (residual = -2.44), suggesting meaningful underlying differences in problem-solving strategies across clusters. Indicating that the answer choice a student picks is not independent of their cluster. Certain clusters are more likely to cluster specific answers than we’d expect by random chance. These patterns support the interpretation of Cluster 1 as representing a correct strategy, and Cluster 3 as a Wrong Answer 1 aligned group. While Cluster 2 has no strong allegiance (residual scores primarily less than $|2|$), supporting them behaving as ambiguous clusters (evenly distributed answers into the cluster). Cluster 4 has shown to favor Wrong Answer 2 (residual = +4.9) and not favor the Correct Answer (residual = -3.13). Based on the representative essay, it was hard to classify Cluster 4; however, we found it likely to either favor Wrong Answer 2 or 3. This result indicates that the cluster favors Wrong Answer 2 more than by chance.

\subsection{Limitations} 

This method has not been tested to the extremes. Pushing it to identify what datasets it works best on is necessary, along with adjusting parameters. This clustering algorithm will likely work best with a diverse dataset, i.e., student responses to a broad question. With a diverse dataset, the number of extracted clusters can be increased which will likely lead to discovering a unknown patterns. There was just enough variance in this dataset to see distinct clusters that correspond to some of the answers. We aim to apply this method to different types of dataset in the coming months. Engaging in this exercise will reveal the limits of this method and ways to improve the process. 

Another limitation with this study is that we do not have any human scored strategy essays to compare with the clustered results. To address this in the future, we could code a small, random subset of student strategies, which would add valuable insights into comparing this method to humans. 

\subsection{Implications and Future Work}

This clustering method has implications for PER. It can be used to understand the general ideas expressed by students as a class. It can also be used to track how those ideas change from week to week or semester to semester. This method has potential for revealing hidden patterns that humans miss; in addition, to quickly measuring the overall themes of a class of student responses. We encourage other researchers to start exploring new methods such as this with their own datasets. These new tools open a whole world of possibilities for exploration into large datasets of student written responses. 

\section{Conclusions}
Overall, the hybrid clustering approach resulted in more distinct clusters. This approach outperforms K-Means and HDBSCAN in terms of evaluation metrics and well-defined clusters illustrated by the scatter plots, silhouette scores, and Davies-Bouldin Index. In addition, compared directly with the student responses, it has a statistically significant result, Figure \ref{fig:distribution}. For the answers that a large portion of the students chose, we see a statistically significant matching with the expected cluster. In conjunction with clearer clusters, we were able to see statistically significant results in essays that were clustered in the "correct cluster" and students whom selected the correct answer.

The goal of this project was not to develop a method to grade student responses. However, this method could be used to extract the relevant strategies students primarily when attempting a multiple-choice question. While not reliable for grading student responses, this method can be used by educators to see relevant or possibly unseen patterns quickly and efficiently.

\section{Acknowledgments}

This research is supported in part by the U.S. National Science Foundation grants 2111138 and 2300648. Any opinions, results, and findings expressed here are those of the authors and not of the Foundation. 

\bibliography{references}

\begin{thebibliography}{35}%
\makeatletter
\providecommand \@ifxundefined [1]{%
 \@ifx{#1\undefined}
}%
\providecommand \@ifnum [1]{%
 \ifnum #1\expandafter \@firstoftwo
 \else \expandafter \@secondoftwo
 \fi
}%
\providecommand \@ifx [1]{%
 \ifx #1\expandafter \@firstoftwo
 \else \expandafter \@secondoftwo
 \fi
}%
\providecommand \natexlab [1]{#1}%
\providecommand \enquote  [1]{``#1''}%
\providecommand \bibnamefont  [1]{#1}%
\providecommand \bibfnamefont [1]{#1}%
\providecommand \citenamefont [1]{#1}%
\providecommand \href@noop [0]{\@secondoftwo}%
\providecommand \href [0]{\begingroup \@sanitize@url \@href}%
\providecommand \@href[1]{\@@startlink{#1}\@@href}%
\providecommand \@@href[1]{\endgroup#1\@@endlink}%
\providecommand \@sanitize@url [0]{\catcode `\\12\catcode `\$12\catcode `\&12\catcode `\#12\catcode `\^12\catcode `\_12\catcode `\%12\relax}%
\providecommand \@@startlink[1]{}%
\providecommand \@@endlink[0]{}%
\providecommand \url  [0]{\begingroup\@sanitize@url \@url }%
\providecommand \@url [1]{\endgroup\@href {#1}{\urlprefix }}%
\providecommand \urlprefix  [0]{URL }%
\providecommand \Eprint [0]{\href }%
\providecommand \doibase [0]{https://doi.org/}%
\providecommand \selectlanguage [0]{\@gobble}%
\providecommand \bibinfo  [0]{\@secondoftwo}%
\providecommand \bibfield  [0]{\@secondoftwo}%
\providecommand \translation [1]{[#1]}%
\providecommand \BibitemOpen [0]{}%
\providecommand \bibitemStop [0]{}%
\providecommand \bibitemNoStop [0]{.\EOS\space}%
\providecommand \EOS [0]{\spacefactor3000\relax}%
\providecommand \BibitemShut  [1]{\csname bibitem#1\endcsname}%
\let\auto@bib@innerbib\@empty
\bibitem [{\citenamefont {MacQueen}\ \emph {et~al.}(1967)\citenamefont {MacQueen} \emph {et~al.}}]{macqueen1967some}%
  \BibitemOpen
  \bibfield  {author} {\bibinfo {author} {\bibfnamefont {J.}~\bibnamefont {MacQueen}} \emph {et~al.},\ }\bibfield  {title} {\bibinfo {title} {Some methods for classification and analysis of multivariate observations},\ }in\ \href@noop {} {\emph {\bibinfo {booktitle} {Proceedings of the fifth Berkeley symposium on mathematical statistics and probability}}},\ Vol.~\bibinfo {volume} {1}\ (\bibinfo {organization} {Oakland, CA, USA},\ \bibinfo {year} {1967})\ pp.\ \bibinfo {pages} {281--297}\BibitemShut {NoStop}%
\bibitem [{\citenamefont {McInnes}\ \emph {et~al.}(2017)\citenamefont {McInnes}, \citenamefont {Healy}, \citenamefont {Astels} \emph {et~al.}}]{mcinnes2017hdbscan}%
  \BibitemOpen
  \bibfield  {author} {\bibinfo {author} {\bibfnamefont {L.}~\bibnamefont {McInnes}}, \bibinfo {author} {\bibfnamefont {J.}~\bibnamefont {Healy}}, \bibinfo {author} {\bibfnamefont {S.}~\bibnamefont {Astels}}, \emph {et~al.},\ }\bibfield  {title} {\bibinfo {title} {hdbscan: Hierarchical density based clustering.},\ }\href@noop {} {\bibfield  {journal} {\bibinfo  {journal} {J. Open Source Softw.}\ }\textbf {\bibinfo {volume} {2}},\ \bibinfo {pages} {205} (\bibinfo {year} {2017})}\BibitemShut {NoStop}%
\bibitem [{\citenamefont {Rousseeuw}(1987)}]{rousseeuw1987silhouettes}%
  \BibitemOpen
  \bibfield  {author} {\bibinfo {author} {\bibfnamefont {P.~J.}\ \bibnamefont {Rousseeuw}},\ }\bibfield  {title} {\bibinfo {title} {Silhouettes: a graphical aid to the interpretation and validation of cluster analysis},\ }\href@noop {} {\bibfield  {journal} {\bibinfo  {journal} {Journal of computational and applied mathematics}\ }\textbf {\bibinfo {volume} {20}},\ \bibinfo {pages} {53} (\bibinfo {year} {1987})}\BibitemShut {NoStop}%
\bibitem [{\citenamefont {Davies}\ and\ \citenamefont {Bouldin}(1979)}]{davies1979cluster}%
  \BibitemOpen
  \bibfield  {author} {\bibinfo {author} {\bibfnamefont {D.~L.}\ \bibnamefont {Davies}}\ and\ \bibinfo {author} {\bibfnamefont {D.~W.}\ \bibnamefont {Bouldin}},\ }\bibfield  {title} {\bibinfo {title} {A cluster separation measure},\ }\href@noop {} {\bibfield  {journal} {\bibinfo  {journal} {IEEE transactions on pattern analysis and machine intelligence}\ ,\ \bibinfo {pages} {224}} (\bibinfo {year} {1979})}\BibitemShut {NoStop}%
\bibitem [{\citenamefont {States}(2013)}]{ngss2013next}%
  \BibitemOpen
  \bibfield  {author} {\bibinfo {author} {\bibfnamefont {N.~L.}\ \bibnamefont {States}},\ }\href@noop {} {\emph {\bibinfo {title} {Next generation science standards: For states, by states}}}\ (\bibinfo  {publisher} {National Academies Press},\ \bibinfo {year} {2013})\BibitemShut {NoStop}%
\bibitem [{\citenamefont {Leonard}\ \emph {et~al.}(1996)\citenamefont {Leonard}, \citenamefont {Dufresne},\ and\ \citenamefont {Mestre}}]{leonard1996using}%
  \BibitemOpen
  \bibfield  {author} {\bibinfo {author} {\bibfnamefont {W.~J.}\ \bibnamefont {Leonard}}, \bibinfo {author} {\bibfnamefont {R.~J.}\ \bibnamefont {Dufresne}},\ and\ \bibinfo {author} {\bibfnamefont {J.~P.}\ \bibnamefont {Mestre}},\ }\bibfield  {title} {\bibinfo {title} {Using qualitative problem-solving strategies to highlight the role of conceptual knowledge in solving problems},\ }\href@noop {} {\bibfield  {journal} {\bibinfo  {journal} {American Journal of Physics}\ }\textbf {\bibinfo {volume} {64}},\ \bibinfo {pages} {1495} (\bibinfo {year} {1996})}\BibitemShut {NoStop}%
\bibitem [{\citenamefont {Mestre}\ \emph {et~al.}(1997)\citenamefont {Mestre}, \citenamefont {Gerace}, \citenamefont {Dufresne},\ and\ \citenamefont {Leonard}}]{mestre1997promoting}%
  \BibitemOpen
  \bibfield  {author} {\bibinfo {author} {\bibfnamefont {J.~P.}\ \bibnamefont {Mestre}}, \bibinfo {author} {\bibfnamefont {W.~J.}\ \bibnamefont {Gerace}}, \bibinfo {author} {\bibfnamefont {R.~J.}\ \bibnamefont {Dufresne}},\ and\ \bibinfo {author} {\bibfnamefont {W.~J.}\ \bibnamefont {Leonard}},\ }\bibfield  {title} {\bibinfo {title} {Promoting active learning in large classes using a classroom communication system},\ }in\ \href@noop {} {\emph {\bibinfo {booktitle} {AIP Conference Proceedings}}},\ Vol.\ \bibinfo {volume} {399}\ (\bibinfo {organization} {American Institute of Physics},\ \bibinfo {year} {1997})\ pp.\ \bibinfo {pages} {1019--1036}\BibitemShut {NoStop}%
\bibitem [{\citenamefont {Maloney}\ \emph {et~al.}(2011)\citenamefont {Maloney}, \citenamefont {Ansari},\ and\ \citenamefont {Fugelsang}}]{maloney2011rapid}%
  \BibitemOpen
  \bibfield  {author} {\bibinfo {author} {\bibfnamefont {E.~A.}\ \bibnamefont {Maloney}}, \bibinfo {author} {\bibfnamefont {D.}~\bibnamefont {Ansari}},\ and\ \bibinfo {author} {\bibfnamefont {J.~A.}\ \bibnamefont {Fugelsang}},\ }\bibfield  {title} {\bibinfo {title} {Rapid communication: The effect of mathematics anxiety on the processing of numerical magnitude},\ }\href@noop {} {\bibfield  {journal} {\bibinfo  {journal} {Quarterly Journal of Experimental Psychology}\ }\textbf {\bibinfo {volume} {64}},\ \bibinfo {pages} {10} (\bibinfo {year} {2011})}\BibitemShut {NoStop}%
\bibitem [{\citenamefont {Dufresne}\ \emph {et~al.}(1992)\citenamefont {Dufresne}, \citenamefont {Gerace}, \citenamefont {Hardiman},\ and\ \citenamefont {Mestre}}]{dufresne1992constraining}%
  \BibitemOpen
  \bibfield  {author} {\bibinfo {author} {\bibfnamefont {R.~J.}\ \bibnamefont {Dufresne}}, \bibinfo {author} {\bibfnamefont {W.~J.}\ \bibnamefont {Gerace}}, \bibinfo {author} {\bibfnamefont {P.~T.}\ \bibnamefont {Hardiman}},\ and\ \bibinfo {author} {\bibfnamefont {J.~P.}\ \bibnamefont {Mestre}},\ }\bibfield  {title} {\bibinfo {title} {Constraining novices to perform expertlike problem analyses: Effects on schema acquisition},\ }\href@noop {} {\bibfield  {journal} {\bibinfo  {journal} {The Journal of the Learning Sciences}\ }\textbf {\bibinfo {volume} {2}},\ \bibinfo {pages} {307} (\bibinfo {year} {1992})}\BibitemShut {NoStop}%
\bibitem [{\citenamefont {Gerace}\ \emph {et~al.}(2001)\citenamefont {Gerace}, \citenamefont {Dufresne}, \citenamefont {Leonard},\ and\ \citenamefont {Mestre}}]{gerace2001problem}%
  \BibitemOpen
  \bibfield  {author} {\bibinfo {author} {\bibfnamefont {W.~J.}\ \bibnamefont {Gerace}}, \bibinfo {author} {\bibfnamefont {R.}~\bibnamefont {Dufresne}}, \bibinfo {author} {\bibfnamefont {W.}~\bibnamefont {Leonard}},\ and\ \bibinfo {author} {\bibfnamefont {J.}~\bibnamefont {Mestre}},\ }\bibfield  {title} {\bibinfo {title} {Problem solving and conceptual understanding},\ }in\ \href@noop {} {\emph {\bibinfo {booktitle} {Proceedings of the 2001 Physics education research conference}}}\ (\bibinfo {organization} {Physics Education Research Conference, Annual Conference},\ \bibinfo {year} {2001})\ p.~\bibinfo {pages} {33}\BibitemShut {NoStop}%
\bibitem [{\citenamefont {Mestre}\ \emph {et~al.}(1993)\citenamefont {Mestre}, \citenamefont {Dufresne}, \citenamefont {Gerace}, \citenamefont {Hardiman},\ and\ \citenamefont {Touger}}]{mestre1993promoting}%
  \BibitemOpen
  \bibfield  {author} {\bibinfo {author} {\bibfnamefont {J.~P.}\ \bibnamefont {Mestre}}, \bibinfo {author} {\bibfnamefont {R.~J.}\ \bibnamefont {Dufresne}}, \bibinfo {author} {\bibfnamefont {W.~J.}\ \bibnamefont {Gerace}}, \bibinfo {author} {\bibfnamefont {P.~T.}\ \bibnamefont {Hardiman}},\ and\ \bibinfo {author} {\bibfnamefont {J.~S.}\ \bibnamefont {Touger}},\ }\bibfield  {title} {\bibinfo {title} {Promoting skilled problem-solving behavior among beginning physics students},\ }\href@noop {} {\bibfield  {journal} {\bibinfo  {journal} {Journal of research in science teaching}\ }\textbf {\bibinfo {volume} {30}},\ \bibinfo {pages} {303} (\bibinfo {year} {1993})}\BibitemShut {NoStop}%
\bibitem [{\citenamefont {Docktor}\ \emph {et~al.}(2010)\citenamefont {Docktor}, \citenamefont {Strand}, \citenamefont {Mestre},\ and\ \citenamefont {Ross}}]{docktor2010conceptual}%
  \BibitemOpen
  \bibfield  {author} {\bibinfo {author} {\bibfnamefont {J.~L.}\ \bibnamefont {Docktor}}, \bibinfo {author} {\bibfnamefont {N.~E.}\ \bibnamefont {Strand}}, \bibinfo {author} {\bibfnamefont {J.~P.}\ \bibnamefont {Mestre}},\ and\ \bibinfo {author} {\bibfnamefont {B.~H.}\ \bibnamefont {Ross}},\ }\bibfield  {title} {\bibinfo {title} {A conceptual approach to physics problem solving},\ }in\ \href@noop {} {\emph {\bibinfo {booktitle} {AIP Conference Proceedings}}},\ Vol.\ \bibinfo {volume} {1289}\ (\bibinfo {organization} {American Institute of Physics},\ \bibinfo {year} {2010})\ pp.\ \bibinfo {pages} {137--140}\BibitemShut {NoStop}%
\bibitem [{\citenamefont {Martin}\ \emph {et~al.}(2023)\citenamefont {Martin}, \citenamefont {Kranz}, \citenamefont {Wulff},\ and\ \citenamefont {Graulich}}]{martin2023exploring}%
  \BibitemOpen
  \bibfield  {author} {\bibinfo {author} {\bibfnamefont {P.~P.}\ \bibnamefont {Martin}}, \bibinfo {author} {\bibfnamefont {D.}~\bibnamefont {Kranz}}, \bibinfo {author} {\bibfnamefont {P.}~\bibnamefont {Wulff}},\ and\ \bibinfo {author} {\bibfnamefont {N.}~\bibnamefont {Graulich}},\ }\bibfield  {title} {\bibinfo {title} {Exploring new depths: Applying machine learning for the analysis of student argumentation in chemistry},\ }\href@noop {} {\bibfield  {journal} {\bibinfo  {journal} {Journal of Research in Science Teaching}\ } (\bibinfo {year} {2023})}\BibitemShut {NoStop}%
\bibitem [{\citenamefont {Tschisgale}\ \emph {et~al.}(2023)\citenamefont {Tschisgale}, \citenamefont {Wulff},\ and\ \citenamefont {Kubsch}}]{tschisgale2023}%
  \BibitemOpen
  \bibfield  {author} {\bibinfo {author} {\bibfnamefont {P.}~\bibnamefont {Tschisgale}}, \bibinfo {author} {\bibfnamefont {P.}~\bibnamefont {Wulff}},\ and\ \bibinfo {author} {\bibfnamefont {M.}~\bibnamefont {Kubsch}},\ }\bibfield  {title} {\bibinfo {title} {Integrating artificial intelligence-based methods into qualitative research in physics education research: A case for computational grounded theory},\ }\href@noop {} {\bibfield  {journal} {\bibinfo  {journal} {Physical Review Physics Education Research}\ }\textbf {\bibinfo {volume} {19}},\ \bibinfo {pages} {020123} (\bibinfo {year} {2023})}\BibitemShut {NoStop}%
\bibitem [{\citenamefont {Nelson}(2020)}]{nelson2020}%
  \BibitemOpen
  \bibfield  {author} {\bibinfo {author} {\bibfnamefont {L.~K.}\ \bibnamefont {Nelson}},\ }\bibfield  {title} {\bibinfo {title} {Computational grounded theory: A methodological framework},\ }\href@noop {} {\bibfield  {journal} {\bibinfo  {journal} {Sociological Methods \& Research}\ }\textbf {\bibinfo {volume} {49}},\ \bibinfo {pages} {3} (\bibinfo {year} {2020})}\BibitemShut {NoStop}%
\bibitem [{\citenamefont {Hastie}\ \emph {et~al.}(2009{\natexlab{a}})\citenamefont {Hastie}, \citenamefont {Tibshirani}, \citenamefont {Friedman}, \citenamefont {Hastie}, \citenamefont {Tibshirani},\ and\ \citenamefont {Friedman}}]{hastie2009supervised}%
  \BibitemOpen
  \bibfield  {author} {\bibinfo {author} {\bibfnamefont {T.}~\bibnamefont {Hastie}}, \bibinfo {author} {\bibfnamefont {R.}~\bibnamefont {Tibshirani}}, \bibinfo {author} {\bibfnamefont {J.}~\bibnamefont {Friedman}}, \bibinfo {author} {\bibfnamefont {T.}~\bibnamefont {Hastie}}, \bibinfo {author} {\bibfnamefont {R.}~\bibnamefont {Tibshirani}},\ and\ \bibinfo {author} {\bibfnamefont {J.}~\bibnamefont {Friedman}},\ }\bibfield  {title} {\bibinfo {title} {Overview of supervised learning},\ }\href@noop {} {\bibfield  {journal} {\bibinfo  {journal} {The elements of statistical learning: Data mining, inference, and prediction}\ ,\ \bibinfo {pages} {9}} (\bibinfo {year} {2009}{\natexlab{a}})}\BibitemShut {NoStop}%
\bibitem [{\citenamefont {Zhu}(2005)}]{zhu2005semi}%
  \BibitemOpen
  \bibfield  {author} {\bibinfo {author} {\bibfnamefont {X.~J.}\ \bibnamefont {Zhu}},\ }\href {https://minds.wisconsin.edu/handle/1793/60444} {\emph {\bibinfo {title} {Semi-Supervised Learning Literature Survey}}},\ \bibinfo {type} {Technical Report}\ \bibinfo {number} {TR1530}\ (\bibinfo  {institution} {University of Wisconsin-Madison, Department of Computer Sciences},\ \bibinfo {year} {2005})\BibitemShut {NoStop}%
\bibitem [{\citenamefont {Hastie}\ \emph {et~al.}(2009{\natexlab{b}})\citenamefont {Hastie}, \citenamefont {Tibshirani}, \citenamefont {Friedman}, \citenamefont {Hastie}, \citenamefont {Tibshirani},\ and\ \citenamefont {Friedman}}]{hastie2009unsupervised}%
  \BibitemOpen
  \bibfield  {author} {\bibinfo {author} {\bibfnamefont {T.}~\bibnamefont {Hastie}}, \bibinfo {author} {\bibfnamefont {R.}~\bibnamefont {Tibshirani}}, \bibinfo {author} {\bibfnamefont {J.}~\bibnamefont {Friedman}}, \bibinfo {author} {\bibfnamefont {T.}~\bibnamefont {Hastie}}, \bibinfo {author} {\bibfnamefont {R.}~\bibnamefont {Tibshirani}},\ and\ \bibinfo {author} {\bibfnamefont {J.}~\bibnamefont {Friedman}},\ }\bibfield  {title} {\bibinfo {title} {Unsupervised learning},\ }\href@noop {} {\bibfield  {journal} {\bibinfo  {journal} {The elements of statistical learning: Data mining, inference, and prediction}\ ,\ \bibinfo {pages} {485}} (\bibinfo {year} {2009}{\natexlab{b}})}\BibitemShut {NoStop}%
\bibitem [{\citenamefont {Ester}\ \emph {et~al.}(1996)\citenamefont {Ester}, \citenamefont {Kriegel}, \citenamefont {Sander}, \citenamefont {Xu} \emph {et~al.}}]{ester1996density}%
  \BibitemOpen
  \bibfield  {author} {\bibinfo {author} {\bibfnamefont {M.}~\bibnamefont {Ester}}, \bibinfo {author} {\bibfnamefont {H.-P.}\ \bibnamefont {Kriegel}}, \bibinfo {author} {\bibfnamefont {J.}~\bibnamefont {Sander}}, \bibinfo {author} {\bibfnamefont {X.}~\bibnamefont {Xu}}, \emph {et~al.},\ }\bibfield  {title} {\bibinfo {title} {A density-based algorithm for discovering clusters in large spatial databases with noise},\ }in\ \href@noop {} {\emph {\bibinfo {booktitle} {kdd}}},\ Vol.~\bibinfo {volume} {96}\ (\bibinfo {year} {1996})\ pp.\ \bibinfo {pages} {226--231}\BibitemShut {NoStop}%
\bibitem [{\citenamefont {Campello}\ \emph {et~al.}(2015)\citenamefont {Campello}, \citenamefont {Moulavi}, \citenamefont {Zimek},\ and\ \citenamefont {Sander}}]{campello2015}%
  \BibitemOpen
  \bibfield  {author} {\bibinfo {author} {\bibfnamefont {R.~J. G.~B.}\ \bibnamefont {Campello}}, \bibinfo {author} {\bibfnamefont {D.}~\bibnamefont {Moulavi}}, \bibinfo {author} {\bibfnamefont {A.}~\bibnamefont {Zimek}},\ and\ \bibinfo {author} {\bibfnamefont {J.}~\bibnamefont {Sander}},\ }\bibfield  {title} {\bibinfo {title} {Hierarchical density estimates for data clustering, visualization, and outlier detection},\ }\bibfield  {journal} {\bibinfo  {journal} {ACM Trans. Knowl. Discov. Data}\ }\textbf {\bibinfo {volume} {10}},\ \href {https://doi.org/10.1145/2733381} {10.1145/2733381} (\bibinfo {year} {2015})\BibitemShut {NoStop}%
\bibitem [{\citenamefont {Docktor}\ and\ \citenamefont {Mestre}(2014)}]{doktor2014}%
  \BibitemOpen
  \bibfield  {author} {\bibinfo {author} {\bibfnamefont {J.~L.}\ \bibnamefont {Docktor}}\ and\ \bibinfo {author} {\bibfnamefont {J.~P.}\ \bibnamefont {Mestre}},\ }\bibfield  {title} {\bibinfo {title} {Synthesis of discipline-based education research in physics},\ }\href {https://doi.org/10.1103/PhysRevSTPER.10.020119} {\bibfield  {journal} {\bibinfo  {journal} {Phys. Rev. ST Phys. Educ. Res.}\ }\textbf {\bibinfo {volume} {10}},\ \bibinfo {pages} {020119} (\bibinfo {year} {2014})}\BibitemShut {NoStop}%
\bibitem [{\citenamefont {Tuminaro}\ and\ \citenamefont {Redish}(2007)}]{Tulminaro2007}%
  \BibitemOpen
  \bibfield  {author} {\bibinfo {author} {\bibfnamefont {J.}~\bibnamefont {Tuminaro}}\ and\ \bibinfo {author} {\bibfnamefont {E.~F.}\ \bibnamefont {Redish}},\ }\bibfield  {title} {\bibinfo {title} {Elements of a cognitive model of physics problem solving: Epistemic games},\ }\href {https://doi.org/10.1103/PhysRevSTPER.3.020101} {\bibfield  {journal} {\bibinfo  {journal} {Phys. Rev. ST Phys. Educ. Res.}\ }\textbf {\bibinfo {volume} {3}},\ \bibinfo {pages} {020101} (\bibinfo {year} {2007})}\BibitemShut {NoStop}%
\bibitem [{\citenamefont {Van~Heuvelen}(1991)}]{van1991learning}%
  \BibitemOpen
  \bibfield  {author} {\bibinfo {author} {\bibfnamefont {A.}~\bibnamefont {Van~Heuvelen}},\ }\bibfield  {title} {\bibinfo {title} {Learning to think like a physicist: A review of research-based instructional strategies},\ }\href@noop {} {\bibfield  {journal} {\bibinfo  {journal} {American Journal of physics}\ }\textbf {\bibinfo {volume} {59}},\ \bibinfo {pages} {891} (\bibinfo {year} {1991})}\BibitemShut {NoStop}%
\bibitem [{\citenamefont {Dufresne}\ \emph {et~al.}(1997)\citenamefont {Dufresne}, \citenamefont {Gerace},\ and\ \citenamefont {Leonard}}]{dufresne1997solving}%
  \BibitemOpen
  \bibfield  {author} {\bibinfo {author} {\bibfnamefont {R.~J.}\ \bibnamefont {Dufresne}}, \bibinfo {author} {\bibfnamefont {W.~J.}\ \bibnamefont {Gerace}},\ and\ \bibinfo {author} {\bibfnamefont {W.~J.}\ \bibnamefont {Leonard}},\ }\bibfield  {title} {\bibinfo {title} {Solving physics problems with multiple representations},\ }\href@noop {} {\bibfield  {journal} {\bibinfo  {journal} {Physics Teacher}\ }\textbf {\bibinfo {volume} {35}},\ \bibinfo {pages} {270} (\bibinfo {year} {1997})}\BibitemShut {NoStop}%
\bibitem [{\citenamefont {Caramazza}\ \emph {et~al.}(1981)\citenamefont {Caramazza}, \citenamefont {McCloskey},\ and\ \citenamefont {Green}}]{caramazza1981naive}%
  \BibitemOpen
  \bibfield  {author} {\bibinfo {author} {\bibfnamefont {A.}~\bibnamefont {Caramazza}}, \bibinfo {author} {\bibfnamefont {M.}~\bibnamefont {McCloskey}},\ and\ \bibinfo {author} {\bibfnamefont {B.}~\bibnamefont {Green}},\ }\bibfield  {title} {\bibinfo {title} {Naive beliefs in “sophisticated” subjects: Misconceptions about trajectories of objects},\ }\href@noop {} {\bibfield  {journal} {\bibinfo  {journal} {Cognition}\ }\textbf {\bibinfo {volume} {9}},\ \bibinfo {pages} {117} (\bibinfo {year} {1981})}\BibitemShut {NoStop}%
\bibitem [{\citenamefont {Eylon}\ and\ \citenamefont {Reif}(1984)}]{eylon1984effects}%
  \BibitemOpen
  \bibfield  {author} {\bibinfo {author} {\bibfnamefont {B.-S.}\ \bibnamefont {Eylon}}\ and\ \bibinfo {author} {\bibfnamefont {F.}~\bibnamefont {Reif}},\ }\bibfield  {title} {\bibinfo {title} {Effects of knowledge organization on task performance},\ }\href@noop {} {\bibfield  {journal} {\bibinfo  {journal} {Cognition and instruction}\ }\textbf {\bibinfo {volume} {1}},\ \bibinfo {pages} {5} (\bibinfo {year} {1984})}\BibitemShut {NoStop}%
\bibitem [{\citenamefont {Larkin}\ \emph {et~al.}(1980)\citenamefont {Larkin}, \citenamefont {McDermott}, \citenamefont {Simon},\ and\ \citenamefont {Simon}}]{larkin1980expert}%
  \BibitemOpen
  \bibfield  {author} {\bibinfo {author} {\bibfnamefont {J.}~\bibnamefont {Larkin}}, \bibinfo {author} {\bibfnamefont {J.}~\bibnamefont {McDermott}}, \bibinfo {author} {\bibfnamefont {D.~P.}\ \bibnamefont {Simon}},\ and\ \bibinfo {author} {\bibfnamefont {H.~A.}\ \bibnamefont {Simon}},\ }\bibfield  {title} {\bibinfo {title} {Expert and novice performance in solving physics problems},\ }\href@noop {} {\bibfield  {journal} {\bibinfo  {journal} {Science}\ }\textbf {\bibinfo {volume} {208}},\ \bibinfo {pages} {1335} (\bibinfo {year} {1980})}\BibitemShut {NoStop}%
\bibitem [{\citenamefont {Hsu}\ \emph {et~al.}(2004)\citenamefont {Hsu}, \citenamefont {Brewe}, \citenamefont {Foster},\ and\ \citenamefont {Harper}}]{hsu2004resource}%
  \BibitemOpen
  \bibfield  {author} {\bibinfo {author} {\bibfnamefont {L.}~\bibnamefont {Hsu}}, \bibinfo {author} {\bibfnamefont {E.}~\bibnamefont {Brewe}}, \bibinfo {author} {\bibfnamefont {T.~M.}\ \bibnamefont {Foster}},\ and\ \bibinfo {author} {\bibfnamefont {K.~A.}\ \bibnamefont {Harper}},\ }\bibfield  {title} {\bibinfo {title} {Resource letter rps-1: Research in problem solving},\ }\href@noop {} {\bibfield  {journal} {\bibinfo  {journal} {American journal of physics}\ }\textbf {\bibinfo {volume} {72}},\ \bibinfo {pages} {1147} (\bibinfo {year} {2004})}\BibitemShut {NoStop}%
\bibitem [{\citenamefont {and}(1987)}]{Hartigan01031987}%
  \BibitemOpen
  \bibfield  {author} {\bibinfo {author} {\bibfnamefont {J.~A.~H.}\ \bibnamefont {and}},\ }\bibfield  {title} {\bibinfo {title} {Estimation of a convex density contour in two dimensions},\ }\href {https://doi.org/10.1080/01621459.1987.10478428} {\bibfield  {journal} {\bibinfo  {journal} {Journal of the American Statistical Association}\ }\textbf {\bibinfo {volume} {82}},\ \bibinfo {pages} {267} (\bibinfo {year} {1987})},\ \Eprint {https://arxiv.org/abs/https://www.tandfonline.com/doi/pdf/10.1080/01621459.1987.10478428} {https://www.tandfonline.com/doi/pdf/10.1080/01621459.1987.10478428} \BibitemShut {NoStop}%
\bibitem [{\citenamefont {Chabay}\ and\ \citenamefont {Sherwood}(2011)}]{chabay2011matter}%
  \BibitemOpen
  \bibfield  {author} {\bibinfo {author} {\bibfnamefont {R.}~\bibnamefont {Chabay}}\ and\ \bibinfo {author} {\bibfnamefont {B.}~\bibnamefont {Sherwood}},\ }\href {https://books.google.com/books?id=8oyNPd5QbYgC} {\emph {\bibinfo {title} {Matter and Interactions}}},\ Matter \& Interactions\ (\bibinfo  {publisher} {Wiley},\ \bibinfo {year} {2011})\BibitemShut {NoStop}%
\bibitem [{\citenamefont {Loper}\ and\ \citenamefont {Bird}(2002)}]{loper2002nltk}%
  \BibitemOpen
  \bibfield  {author} {\bibinfo {author} {\bibfnamefont {E.}~\bibnamefont {Loper}}\ and\ \bibinfo {author} {\bibfnamefont {S.}~\bibnamefont {Bird}},\ }\bibfield  {title} {\bibinfo {title} {Nltk: The natural language toolkit},\ }\href@noop {} {\bibfield  {journal} {\bibinfo  {journal} {arXiv preprint cs/0205028}\ } (\bibinfo {year} {2002})}\BibitemShut {NoStop}%
\bibitem [{\citenamefont {Barrus}(2019)}]{barrus2019pure}%
  \BibitemOpen
  \bibfield  {author} {\bibinfo {author} {\bibfnamefont {T.}~\bibnamefont {Barrus}},\ }\href@noop {} {\bibinfo {title} {Pure python spell checker based on work by peter norvig}} (\bibinfo {year} {2019})\BibitemShut {NoStop}%
\bibitem [{\citenamefont {Pedregosa}\ \emph {et~al.}(2011)\citenamefont {Pedregosa}, \citenamefont {Varoquaux}, \citenamefont {Gramfort}, \citenamefont {Michel}, \citenamefont {Thirion}, \citenamefont {Grisel}, \citenamefont {Blondel}, \citenamefont {Prettenhofer}, \citenamefont {Weiss}, \citenamefont {Dubourg} \emph {et~al.}}]{pedregosa2011scikit}%
  \BibitemOpen
  \bibfield  {author} {\bibinfo {author} {\bibfnamefont {F.}~\bibnamefont {Pedregosa}}, \bibinfo {author} {\bibfnamefont {G.}~\bibnamefont {Varoquaux}}, \bibinfo {author} {\bibfnamefont {A.}~\bibnamefont {Gramfort}}, \bibinfo {author} {\bibfnamefont {V.}~\bibnamefont {Michel}}, \bibinfo {author} {\bibfnamefont {B.}~\bibnamefont {Thirion}}, \bibinfo {author} {\bibfnamefont {O.}~\bibnamefont {Grisel}}, \bibinfo {author} {\bibfnamefont {M.}~\bibnamefont {Blondel}}, \bibinfo {author} {\bibfnamefont {P.}~\bibnamefont {Prettenhofer}}, \bibinfo {author} {\bibfnamefont {R.}~\bibnamefont {Weiss}}, \bibinfo {author} {\bibfnamefont {V.}~\bibnamefont {Dubourg}}, \emph {et~al.},\ }\bibfield  {title} {\bibinfo {title} {Scikit-learn: Machine learning in python},\ }\href@noop {} {\bibfield  {journal} {\bibinfo  {journal} {the Journal of machine Learning research}\ }\textbf {\bibinfo {volume} {12}},\ \bibinfo {pages} {2825} (\bibinfo {year} {2011})}\BibitemShut {NoStop}%
\bibitem [{\citenamefont {Van~der Maaten}\ and\ \citenamefont {Hinton}(2008)}]{van2008visualizing}%
  \BibitemOpen
  \bibfield  {author} {\bibinfo {author} {\bibfnamefont {L.}~\bibnamefont {Van~der Maaten}}\ and\ \bibinfo {author} {\bibfnamefont {G.}~\bibnamefont {Hinton}},\ }\bibfield  {title} {\bibinfo {title} {Visualizing data using t-sne.},\ }\href@noop {} {\bibfield  {journal} {\bibinfo  {journal} {Journal of machine learning research}\ }\textbf {\bibinfo {volume} {9}} (\bibinfo {year} {2008})}\BibitemShut {NoStop}%
\bibitem [{\citenamefont {Hunter}(2007)}]{hunter2007matplotlib}%
  \BibitemOpen
  \bibfield  {author} {\bibinfo {author} {\bibfnamefont {J.~D.}\ \bibnamefont {Hunter}},\ }\bibfield  {title} {\bibinfo {title} {Matplotlib: A 2d graphics environment},\ }\href@noop {} {\bibfield  {journal} {\bibinfo  {journal} {Computing in science \& engineering}\ }\textbf {\bibinfo {volume} {9}},\ \bibinfo {pages} {90} (\bibinfo {year} {2007})}\BibitemShut {NoStop}%
\end{thebibliography}%
\clearpage
\appendix
\onecolumngrid

\appendix
\onecolumngrid

\end{document}